# A Full Image of the Wormhole Attacks
## Towards Introducing Complex Wormhole Attacks in wireless Ad Hoc Networks

Marianne Azer
Computer Dept.
National Telecommunication Institute
Cairo, Egypt
marazer@nti.sci.eg

Sherif El-Kassas
Computer Science Dept.
American University in Cairo
Cairo, Egypt
sherif@aucegypt.edu

Magdy El-Soudani
Electronics and Communications Dept.
Faculty of Engineering
Cairo, Egypt
melsoudani@menanet.net

*Abstract*— In this paper, we are concerned of a particularly severe security attack that affects the ad hoc networks routing protocols, it is called the wormhole attack. We can think of wormhole attack as a two phase process launched by one or several malicious nodes. In the first phase, these malicious nodes, called wormhole nodes, try to lure legitimate nodes to send data to other nodes via them. In the second phase, wormhole nodes could exploit the data in variety of ways. We will introduce the wormhole attack modes and classes, and point to its impact and threat on ad hoc networks. We also analyze the wormhole attack modes from an attacker's perspective and suggest new improvements to this type of attacks. We finally summarize and conclude this paper.

*Keywords-attacks; ad hoc network; security; tunneling; wormhole*

## I. INTRODUCTION

With the rapid development in wireless technology, ad hoc networks have emerged in many forms. These networks operate in the license free frequency band and do not require any investment in infrastructure, making them attractive for military and selected commercial applications. However, there are many unsolved problems in ad hoc networks; securing the network being one of the major concerns. Ad hoc networks are vulnerable to attacks due to many reasons; amongst them are the absence of infrastructure, wireless links between nodes, limited physical Protection, and the Lack of a centralized monitoring or management, and the resource constraints. A particularly severe security attack, called the wormhole attack, has been introduced in the context of ad-hoc networks [1], [2], [3]. During the attack [4] a malicious node captures packets from one location in the network, and tunnels them to another malicious node at a distant point, which replays them locally, this is illustrated in Figure 1. The tunnel can be established in many different ways, such as through an out-of-band hidden channel (e.g., a wired link), packet encapsulation, or high powered transmission. This tunnel makes the tunneled packet arrive either sooner or with less number of hops compared to the packets transmitted over normal multihop routes. This creates the illusion that the two end points of the tunnel are very close to each other. In this paper, we explain the attack modes and point to the impact of this attack and its threats. From an attacker's perspective, we analyze each of the attack's modes' benefits and suitable conditions and think how to improve the wormhole attack by introducing the concept of "complex wormhole attacks". The remainder of this paper is organized as follows. Section II introduces the wormhole attack modes, threats, impact on the ad hoc networks applications and routing, and solutions that have been proposed in the literature as a countermeasure for this attack. In section III, we analyze the different attack modes in details, while in section IV we suggest complex wormhole attacks to improve the regular wormhole attacks. Finally, conclusions and future directions are given in section V.

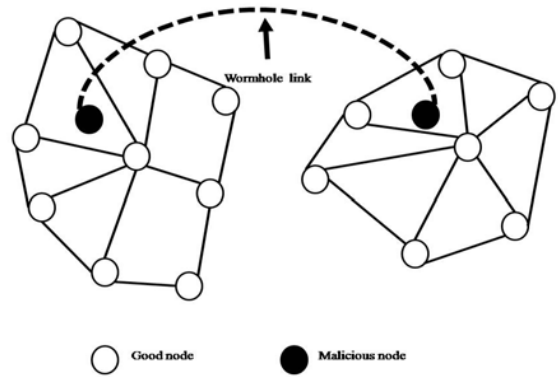

Figure 1. The wormhole attack

## II. THE WORMHOLE ATTACK

In this section we explain the wormhole attacks modes and classes while pointing to the impact of the wormhole attack and the efforts that have been done in the literature to detect and prevent this attack.

*A. Wormhole Attack Modes*

Wormhole attacks can be launched using several modes, among these modes [5], we mention:

**Wormhole using Encapsulation:** In this mode a malicious node at one part of the network and hears the RREQ packet. It tunnels it to a second colluding party at a distant location near the destination. The second party then rebroadcasts the RREQ. The neighbours of the second colluding party receive the RREQ and drop any further legitimate requests that may arrive later on legitimate multihop paths. The result is that the routes between the source and the destination go through the two colluding nodes that will be said to have formed a wormhole between them. This prevents nodes from discovering legitimate paths that are more than two hops away. For example, consider Figure 2 [5] in which nodes *A* and *B* try to discover the shortest path between them, in the presence of the two malicious nodes *X* and *Y*. Node *A* broadcasts a RREQ, *X* gets the RREQ and encapsulates it in a packet destined to *Y* through the path that exists between *X* and *Y* (*U-V-W-Z*). Node *Y* demarshalls the packet, and rebroadcasts it again, which reaches *B*. Note that due to the packet encapsulation, the hop count does not increase during the traversal through *U-V-W-Z*. Concurrently, the R*REQ* travels from *A* to *B* through *C-D-E*. Node *B* now has two routes, the first is four hops long (*A-C-D-E-B*), and the second is apparently three hops long (*A-X-Y-B*). Node *B* will choose the second route since it appears to be the shortest while in reality it is seven hops long. Any routing protocol that uses the metric of shortest path to choose the best route is vulnerable to this mode of wormhole attack.

This mode of the wormhole attack is easy to launch since the two ends of the wormhole do not need to have any cryptographic information, nor do they need any special capabilities, such as a high speed wire line link or a high power source.

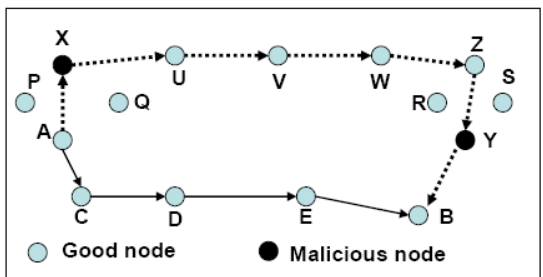

Figure 2. Wormhole through packet encapsulation [5]

**Wormhole using Out-of-Band Channel:** The second mode for this attack is the use of an out of band channel. This channel can be achieved, for example, by using a long-range directional wireless link or a direct wired link. This mode of attack is more difficult to launch than the previous one since it needs specialized hardware capability. Consider the scenario depicted in Figure 3[5]. Node *A* sends a RREQ to node *B*, and nodes *X* and *Y* are malicious nodes having an out-of-band channel between them. Node *X* tunnels the RREQ to *Y*, which is a legitimate neighbor of *B*. Node *Y* broadcasts the packet to its neighbors, including *B*. *B* gets two RREQs—*A-X-Y-B* and *A-C-D-E-F-B*. The first is both shorter and faster than the second, and is thus chosen by *B*.

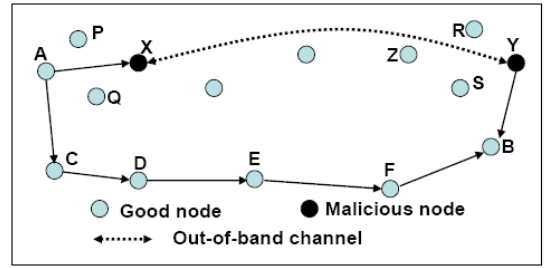

Figure 3. Wormhole through out-of-band channel [5]

**Wormhole with High Power Transmission** Another method is the use of high power transmission. In this mode, when a single malicious node gets a RREQ, it broadcasts the request at a high power level, a capability which is not available to other nodes in the network. Any node that hears the high-power broadcast rebroadcasts it towards the destination. By this method, the malicious node increases its chance to be in the routes established between the source and the destination even without the participation of a colluding node.

**Wormhole using Packet Relay:** Wormhole using Packet Relay is another mode of the wormhole attack in which a malicious node relays packets between two distant nodes to convince them that they are neighbours. It can be launched by even one malicious node. Cooperation by a greater number of malicious nodes serves to expand the neighbour list of a victim node to several hops. It is carried out by an intruder node *X* located within transmission range of legitimate nodes *A* and *B*, where *A* and *B* are not themselves within transmission range of each other. Intruder node *X* merely tunnels control traffic between *A* and *B* (and vice versa), without the modification presumed by the routing protocol e.g. without stating its address as the source in the packets header so that *X* is virtually invisible. This results in an extraneous inexistent *A* - *B* link which in fact is controlled by *X*, as shown in Figure 4. Node *X* can afterwards drop tunneled packets or break this link at will. Two intruder nodes *X* and *X′*, connected by a wireless or wired private medium, can also collude to create a longer (and more harmful) wormhole, as shown in Figure 5. An extraneous *A* - *B* link can be artificially created by an intruder node *X* by wormholing control messages between *A* and *B* as in Figure 4. A longer wormhole can also be created by two colluding intruders *X* and *X′* as in Figure 5. To successfully exploit the wormhole, the attacker must wait until *A* and *B* have exchanged sufficient HELLO messages (through the wormhole) to establish a symmetric link. Until that moment, other tunneled control messages would be rejected, because the OLSR protocol specifies that TC/MID/HNA messages should not be processed if the relayer node (the last hop) is not a symmetric neighbor. However, once created, the *A* - *B* link is at the mercy of the attacker.

*(IJCSIS) International Journal of Computer Science and Information Security, Vol. 1, No. 1, May 2009*

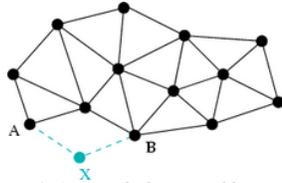

Figure 4. A wormhole created by node *X*

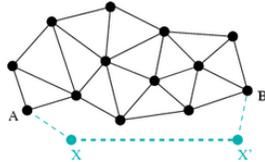

Figure 5. A longer wormhole created by two colluding nodes

**Wormhole using Protocol Deviations:** A wormhole attack can also be done through protocol deviations. During the RREQ forwarding, the nodes typically back off for a random amount of time before forwarding reduce MAC layer collisions. A malicious node can create a wormhole by simply not complying with the protocol and broadcasting without backing off. The purpose is to let the request packet it forwards arrive first at the destination.

The utility of organizing combinations of network attacks as graphs is well established. Network attack graphs represent a collection of possible penetration scenarios in a computer network. The graph can focus on the extent to which an adversary can penetrate a network to achieve a particular goal, given an initial set of capabilities. They represent not only specific attacks but categories of attacks. They can detect previously unseen attacks which have common features with attacks in graphs. Figure 6 depicts an attack graph we have developed for the wormhole attack to illustrate its different modes.

### B. Wormhole Attack Classification

The classification of such an attack facilitates the design of prevention and detection methods. According to whether the attackers are visible on the route, we classify the wormholes into three types: closed, half open, and open [6]. Figure 7 [7] illustrates the three types of wormhole attack.

**Open Wormhole attack:** In this type of wormhole, the attackers include themselves in the RREQ packet header following the route discovery procedure. Other nodes are aware that the malicious nodes lie on the path but they would think that the malicious nodes are direct neighbors.

**Closed Wormhole Attack:** The attackers do not modify the content of the packet, even the packet in a route discovery packet. Instead, they simply tunnel the packet form one side of wormhole to another side and it rebroadcasts the packet.

**Half open wormhole attack:** One side of wormhole does not modify the packet and only another side modifies the packet, following the route discovery procedure.

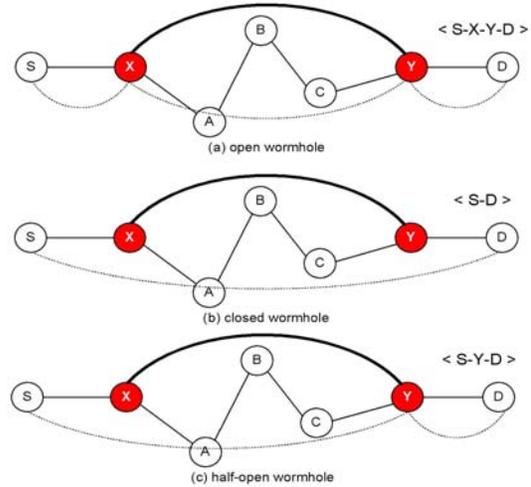

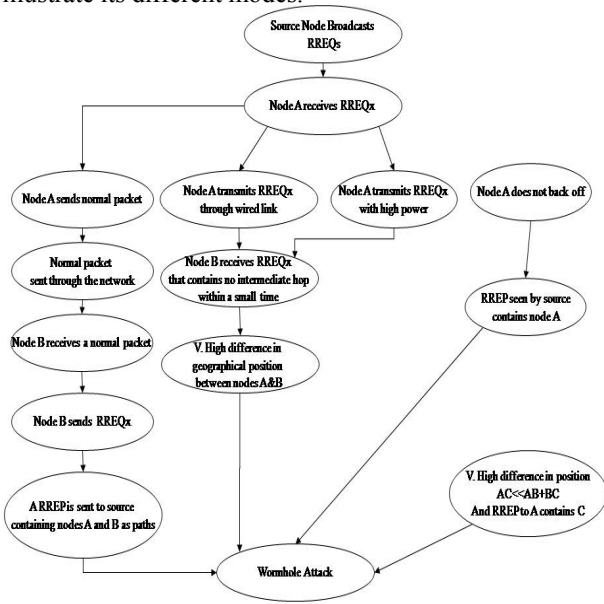

Figure 6. Attack graph of wormhole attack modes

Figure 7. Different types of wormhole attack [7]

### C. Wormhole Attack Threats

A wormhole tunnel could actually be useful if used for forwarding all the packets. However, in its malicious incarnation, it attacks nodes to subvert the correct operation of ad-hoc and sensor network routing protocols. We can think of wormhole attack as a two phase process launched by one or several malicious nodes. In the first phase, the two malicious end points of the tunnel may use it to pass routing traffic to attract routes through them. In the second phase, wormhole nodes could exploit the data in variety of ways. They can disrupt the data flow by selectively dropping or modifying data packets, generating unnecessary routing activities by turning off the wormhole link periodically, etc. The attacker can also simply record the traffic for later analysis. Using wormholes an attacker can also break any protocol that directly or indirectly relies on geographic proximity. For example, target tracking applications in sensor networks can be easily confused in the presence of wormholes. Similarly, wormholes will affect connectivity-based localization algorithms, as two neighbouring nodes are localized nearby and the wormhole links essentially 'fold' the entire network. This can have a major impact as location is a useful service in many protocols and application, and often out-of-band location systems such as GPS are considered expensive or unusable because of the environment. It should be noted that wormholes are dangerous by themselves, even if attackers are diligently forwarding all packets without any disruptions, on some level, providing a communication service to the network. With wormhole in place, affected network nodes do not have a true picture of the network, which may disrupt the localization-based schemes, and hence lead to the wrong decisions, etc. Wormhole can also be used to simply aggregate a large number of network packets for the purpose of traffic analysis or encryption compromise.

The wormhole attack also has a strong impact on both periodic and on demand routing protocols as follows:

**Periodic protocols:** Periodic protocols are based on the distance vector routing algorithm, where each node stores a routing table that contains for each possible destination the associated routing cost, usually in number of hops, and the corresponding next hop towards that destination. Periodically, or when a route change occurs, each node broadcasts its routing table in order to inform its neighbors about possible route changes. Every node that receives a route update adjusts its own routing table based on the broadcast received from the neighboring nodes.

To illustrate the impact of wormhole attacks on periodic protocols, consider Figure 8 which shows an ad hoc network of 13 nodes. A node $n_i$ is connected to a node $n_j$ if the distance between them is less than the communication range $r$. Consider an attacker establishing a wormhole link between nodes $n_9$ and $n_2$, using a low-latency link. When node $n_9$ broadcasts its routing table, node $n_2$ will hear the broadcast via the wormhole and assume it is one hop away from $n_9$. Then, $n_2$ will update its table entries for node $s_9$, reachable via one hop, nodes $\{n_8, n_{10}, n_{11}, n_{12}\}$, reachable via two hops, and broadcast its own routing table. Similarly, the neighbors of $n_2$ will adjust their own routing tables. Note that nodes $\{n_1, n_3, n_4, n_5, n_7\}$ now route via $n_2$ to reach any of the nodes $\{n_9, n_{10}, n_{11}, n_{12}\}$ [8].

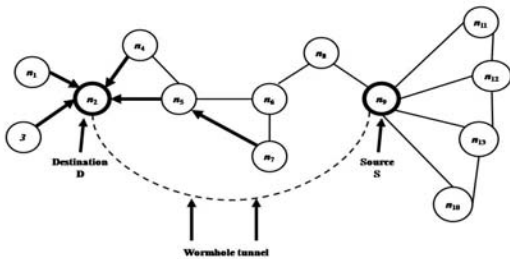

Figure 8. Wormhole attack on a distance vector-based routing protocol

**On-demand protocols:** A wormhole attack against on demand routing protocols can result in similar false route establishment as in the case of periodic protocols. In the route discovery mechanism, a node $S$ initiates a route discovery to node $D$ by broadcasting a RREQ message. All nodes that hear the RREQ message will rebroadcast the request until the destination $D$ has been discovered. Once the destination $D$ is reached, node $D$ will respond with a RREP message. The RREP message will follow a similar route discovery procedure, if the path from $D$ to $S$ has not been previously discovered. If an attacker mounts a wormhole link between the RREQ initiator $S$ and the destination $D$, and if $S, D$ are more than one hop away, then a one-hop route via the wormhole will be established from $S$ to $D$. As an example, consider Figure 9 which is the same topology as in Figure8 Consider that the attacker establishes a wormhole link between nodes $n_9$ and $n_2$ and assume that node $n_9$ wants to send data to node $n_2$. When node $n_9$ broadcasts the RREQ, the attacker will forward the request via the wormhole link to node $n_2$. Node $n_2$ will reply with a RREP and the attacker using wormhole link will forward the reply to node $n_9$. At this point, nodes $n_2$, $n_9$ will establish a route via the wormhole link, as if they were one hop neighbors. Similarly, if any of the nodes $\{n_1, n_3, n_4, n_5, n_7\}$ wants to send data to any of the nodes $\{n_9, n_{10}, n_{11}, n_{12}\}$, the routing paths established will include the wormhole link [8].

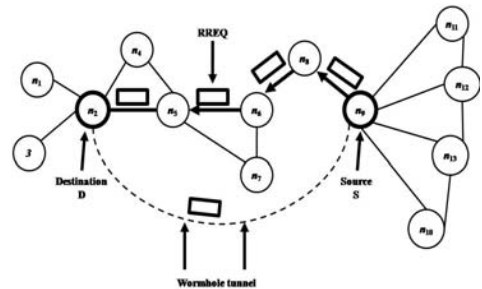

Figure 9. Wormhole attack against an on-demand routing protocol

From the above examples and the existing literature [2], we note that the existence of wormhole links impacts the network routing service performance in the following three ways:

(1) Nodes can become sinkholes [1] without even being aware that they are victims of a wormhole attack (as noted in both Figure 8, and Figure 9, nodes $s_2$, $s_9$ become sinkhole nodes and attract all traffic from surrounding nodes). Hence, a significant amount of traffic is routed through the wormhole link and the attacker can control and observe a significant amount of traffic flow without the need to deploy multiple observation points.

(2) If an attacker kept the wormhole link functional at all times and did not drop any packets, the wormhole would actually provide a useful network service by expediting the packet delivery. However, by selectively dropping packets, the attacker can lower the throughput of the network.

(3) Furthermore, by simply switching the wormhole link on and off, the attacker can trigger a route oscillation within the network, thus leading to a DoS attack, driving the routing service to be unusable [8].

*D. Wormhole Attack Proposed Solutions*

Several solutions have been proposed in the literature for the wormhole attack, the solutions can be categorized into location based, time based, key based, statistics, and graph based solutions. In this section we give a brief overview of these solutions.

**Location and Time Based Solutions**

Most of the proposed wormhole solutions in the literature are based on location or time. Packet leashes have been proposed and specifically two types of packet leashes: geographical and temporal were considered in [2]. The main idea is that by authenticating either an extremely precise timestamp or location information combined with a loose timestamp, a receiver can determine if the packet has traversed an unrealistic distance for the specific network technology used. Packet leashing was added to each packet on each link to restrict the transmission distance of the packet. Two types of packet leashes could be added into the packet. One is geographical leash in which the sender inserts its own position and sending time into the packet, the receiver will estimate the maximum distance between the sender and itself based on its own position and receiving time. If the distance exceeds the transmission range, the packet will be discarded. The other type is temporal leash. This mechanism assumes that the maximum transmission speed of radio signal is the speed of light, thus the expiration time of a packet can be estimated using the maximum transmission range and the speed of light. The expiration time of the packet is inserted into the packet, and then the receiver can check whether the received packet has expired or not based on its receiving time. A drawback of packet leashes is that it requires extremely tight time synchronization and GPS. In [9], secure tracking of node encounters (SECTOR) was proposed. It applied similar principle as packet leashes, with the difference that it measured the distance at a single hop and it required special hardware at each node. Directional antennas can be considered as location based solutions and were used in [3] to prevent the wormhole attack.

A mechanism based on signal strength and geographical information for detecting malicious nodes staging HELLO flood and wormhole attacks was proposed in [10]. The idea is to compare the signal strength of a reception with its expected value, calculated using geographical information and the pre-defined transceiver specification. A protocol for disseminating information about detection of malicious nodes was also proposed.

In [11], another method was suggested in which the sender sets the Destination-only flag such that only the destination can respond to the RREQ packet. Once the RREQ packet reaches the destination, it responds with a RREP with its current position. The sender retrieves the receiver's position from the RREP packet and estimates the lower bound of hops between the sender and the receiver. If the received route is shorter than the estimated shortest path, the corresponding route will be discarded. Otherwise, the sender will select the shortest path corresponding to the estimation. Once a wormhole is detected by the sender, the sender temporarily enables the path with wormhole and sends out a TRACE packet to the receiver. This TRACE packet is forwarded by each intermediate node through the route with wormhole. When a node on the route receives the TRACE packet, it replies the source with its current position and its hop count to the destination. Then, the sender can estimate the increase of hop count at each node using the received position. If the increase of hop count at one node is not one comparing to its previous hop, then this node and its previous hop node are identified as the wormhole. This approach was illustrated with more details in [12], where an end-to-end detection of wormhole attack (EDWA) in wireless ad-hoc networks was explained in details. In addition analysis and simulation results have shown that the end-to-end wormhole detection method is effective when the source and destination are not too far away.

A four way handshaking message exchange method was proposed in [13] for detecting and preventing wormhole attacks against the OLSR routing protocol. To infer suspicious links, new control packets were defined. A timeout value is set for the control packets, and depending on the measured control packets delay, links can be judged as suspicious and then verifies as wormhole links. In [14] an end to end protocol to secure ad hoc networks against wormhole attacks was proposed. The algorithm provides a lower bound on the minimum number of hops on a good route. Any path showing lesser hop-counts is shown to be under attack. The algorithm requires every node to know its location. With very accurate GPS available, this assumption is not unreasonable. Since the protocol does not require speed or time, there is no need for clock synchronization. In the absence of any error in the location, there are no false alarms i.e. no good paths are discarded.

**Key Based Solutions**

For the key based solutions, a scheme was proposed in [15], and [8] depending on location-based keys, a node-to-node authentication scheme, which is not only able to localize the impact of compromised nodes within their vicinity, but also to facilitate the establishment of pairwise keys between neighboring nodes was developed. These schemes only accept messages from authenticated neighbors and discard those messages tunneled from multi-hop-away locations preventing thus the wormhole attack.

**Statistics- Based Solutions**

In [16], a statistical based solution was proposed. The main idea of the proposed scheme SAM was based on the observation that certain statistics of the discovered routes by routing protocols will change dramatically under wormhole attacks. Hence, it was possible to examine such statistics to detect this type of routing attacks and pinpoint the attackers if enough routing information is available (obtained by multi-path routing). Some other schemes use statistical testing to measure the distribution of the number of neighbors or the distance of all pairs of nodes [17].

**Graph-Based Solutions**

One type of wormhole detection involves graph theories. In Multi-Dimensional Scaling Visualization of Wormhole (MDS-VOW) [18] multi-dimensional scaling in graph theory was used to reconstruct the topology of the network. A wormhole attack could cause distortion of network topology which could be detected using graph visualization. In [19], it has been noted that the placement of wormhole influences the network connectivity by creating long links between two sets of nodes located potentially far away. The resulting connectivity graph thus deviates from the true connectivity graph. The detection algorithm essentially looks for forbidden substructures in the connectivity graph that should not be present in a legal connectivity graph. Knowledge of the wireless communication model between the nodes helps the detection algorithm. This is because a communication model can help define what substructures observed in the connectivity graph could be forbidden.

In [20], a graph theoretic framework for modeling wormhole links and deriving the necessary and sufficient conditions for detecting and defending against wormhole attacks was presented. Based on this framework, candidate solution preventing wormholes should construct a communication graph that is a subgraph of the geometric graph defined by the radio range of the network nodes. Making use of our framework, a cryptographic mechanism based on local broadcast keys in order to prevent wormholes was proposed. The solution requires only a small fraction of the nodes to know their location.

**Neighbor-Based solutions**

A wormhole attack prevention algorithm that depends on neighbor monitoring was suggested in [20]. In this method, all nodes monitor their neighbors' behavior when they send RREQ messages to the destination by using a special list called Neighbor List. When a source node receives some RREP messages, it can detect a route under wormhole attack among the routes. Once wormhole node is detected, source node records them in the Wormhole Node List. Even though malicious nodes have been excluded from routing in the past, the nodes have a chance of attack once more. Therefore, the information of wormhole nodes is stored at the source nodes to prevent them taking part in routing again. In [21], another method was suggested. Whenever routing takes place in the network, analysis of the frequencies of links in different routes is done. If any of the links are suspicious, then the available trust information is used to check if the link is that of a wormhole. Following the neighbor monitoring phase, a trust vector of a node containing the trust values to each of its neighbors is calculated. In the trust model used, nodes monitor neighbors based on their packet drop pattern and not on the measure of number of drops. Finally, the algorithm for detection of Wormhole is run during the routing phase. A wormhole attack detection approach based on the probability distribution of the neighboring-node-number, WAPN, which helps the nodes to judge whether a wormhole attack is taking place and whether they are in the influencing area of the attack was proposed in [22]. In [23], wormhole attack defense method was proposed. In the proposed method, each node maintains its neighbors' information. According to the information, each node can identify replayed packet that forwarded by two attackers.

III. ATTACK BENEFITS

In this section, we shall wear an attacker's hat and consider the wormhole attack from an attacker's perspective. To launch a particular wormhole attack mode, an attacker needs to know the way of launching it, its advantages and disadvantages of this attack from his point of view, and its impact of the remaining parts of the network. It is also important to know the challenges that will be faced to launch a successful attack, the suitable number of needed malicious nodes, the suitable attack mode for a specific network topology and the countermeasures that might be used in order to prevent and detect this attack. There are common benefits to all attack modes, which are illusion given to the network nodes that the paths containing malicious nodes have the minimum number of hops or can deliver traffic in a speedier way. To make the comparison clear, all points of comparison are presented in Table 1.

TABLE I. COMPARISON BETWEEN THE WORMHOLE ATTACK MODES FROM AN ATTACKER'S PERSPECTIVE

|  | **Encapsulation** | **Out of Band** | **High Power** | **Packet Relay** | **Protocol Deviations** |
|---|---|---|---|---|---|
| **Attack mode launching method** | Node encapsulates the route request and sends it to colluding node which decapsulates it and forwards the RREQ | Nodes send RREQs between them by using a long-range directional wireless link or a direct wired link | A node gets a RREQ, it broadcasts it at a high power level, Any node that hears the high-power broadcast rebroadcasts it towards the destination | Nodes relay packets between two distant nodes to convince them that they are neighbours | Nodes do not back off to let the request packet it forwards arrive first at the destination. |
| **Advantages** | 1. There is a smaller probability of a RREQ being | 1.Control packet arrives faster since there is no | 1.Control packets arrive faster 2.It has a less | 1. Two nodes think they are neighbors although they are | Control packet arrives faster |
| | discarded than other RREQs that are repeatedly received by intermediate nodes<br><br>2. A RREQ packet arriving at destination does not hold intermediate nodes as hops, and then it appears to have passed through min number of hops. | processing from intermediate nodes<br><br>2.It has a less probability of being discarded than other RREQs that are repeatedly received by intermediate nodes<br><br>3.Control packets arriving at destination, do not hold intermediate nodes as hops, then it appears to have passed through min number of hops. | probability of being discarded than other RREQs that are repeatedly received by intermediate nodes<br>3. Control packet arriving at destination, does not hold intermediate nodes as hops, then it appears to have passed through min number of hops.<br>4.No need for colluding nodes, any node could do the job | not, and every RREQ to be sent to neighbors will arrive to the relaying nodes invisibly.<br>2.Control packet seems to arrive using the minimum number of hops | |
| **Disadvantages** | 1. Resources and time consumption in packet encapsulation and decapsulation | 1. This mode of attack is difficult to launch than the previous one since it needs specialized hardware capability.<br>2. Also the time difference in control packets arrival could be very remarkable. | 1.Needs high power<br>2.Also speed difference could be noticed | 1. Relaying nodes spend resources for processing RREQ packets and hiding their IDs | Does not necessarily provide the minimum number of hops, it is not reliable if collisions occur to give a minimum speed as well |
| **Suitable cases or network topology** | Large number of intermediate nodes, in that case we avoid intermediate processing | A small network so that the speed difference would not be that remarkable | A network with a lot of intermediate nodes between source and destination and wide network range to make a big difference | Victim nodes need to be at least two hops away | Network with large number of nodes to have a big difference from small savings |
| **Challenges to be faced** | 1.Send encapsulated packet to the proper colluding node, while having a predetermined path<br><br>2. If an intermediate node checks the | Need of special hardware and arrangements for out of band channels | Not only enough energy needs to be there, but also power adjustments are needed to make the transmitted RREQ go to some | 1.Insert malicious nodes at proper positions<br>2.Hide malicious node names so that it does not appear on the RREQ packets | Collisions may occur between transmissions of malicious nodes |
| | contents of the sent packet | | suitable neighboring nodes, else the RREQ could go out of the network's range | 3. Choice of optimum number of relaying nodes depending on the victim's distance<br>4. Communication between relaying nodes | |
| **Possible solutions for challenges** | For the predetermined path to be established, colluding nodes could send regular RREQ packets to establish paths.<br><br>For the second challenge of nodes checking packets, complex attacks will solve this problem | | Send a RREQ with different power levels By this, the malicious node will have have a primary network topology. It can then use communication ranges, and number of hops to adjust its power according to the location of the destination | Start by having a large number of relaying nodes and then minimize them to get the optimum performance with a small number of malicious nodes and traffic conquer. Different relaying nodes distribution should also be tried with optimum number | A priority or round robin scheme for malicious nodes packets could be used |
| **Minimum number of malicious nodes** | Two nodes | Two nodes | One node | One node | One node |
| **Impact on other network nodes** | 1. This method attracts all traffic in the wormhole tunnel, thus intermediate nodes within the tunnel can suffer from resource exhaustion due to increase of traffic they have to route<br><br>2. Some nodes might have a wrong picture of the network topology | 1. This method saves the resources of other intermediate nodes that could have been used to process packets, since the packets do not pass by them anyway<br>2. Some nodes might have a wrong picture of the network topology | 1. This method saves the resources of other intermediate nodes that could have been used to process packets, since the packets do not pass by them anyway<br>2. Some nodes might have a wrong picture of the network topology | 1. Victim nodes forward the requests of their neighbors to non neighbor nodes. This could delay or even stop the routing operation in case there are some nodes only surrounded by victim nodes | There is a very high probability of collisions and packets loss |

| **Counter-measures** | Use of statistical methods to get the IDs of repeated nodes and use trace packets to get the proper network topology  Also neighbor monitoring based solutions can help to make sure a RREQ is forwarded | Time based solutions can help as a countermeasure to measure the difference of time | Power measurement equipments could help. Also neighbor monitoring and time based solutions could help | Distance and time based solutions can help | A priority scheme could be used to accept packets in the medium |
|---|---|---|---|---|---|

## IV. WORMHOLE ATTACK SMULATIONS

In this section, we present simulation results obtained using the OPNET modeler. To illustrate the effect of the wormhole attack, we present three scenarios with different number of malicious nodes. The first is launched using only one node, the second used two malicious nodes and the third uses three malicious nodes. In the first two scenarios, the network consists of MANET nodes, and a heavy FTP server sending traffic to three nodes. The AODV routing protocol was configured on all nodes. Ad Hoc Network Parameters and Values for the first two scenarios with one and two malicious nodes contains some of the values assigned to the ad hoc network parameters.

TABLE II. AD HOC NETWORK PARAMETERS AND VALUES FOR THE FIRST TWO SCENARIOS WITH ONE AND TWO MALICIOUS NODES

| Ad Hoc Network Parameters | Values |
|---|---|
| Route Request Retries | 5 |
| Route Request Rate (pkts/sec) | 10 |
| Active Route Time Out (Seconds) | 15 s |
| Hello Interval (Seconds) | Uniform (5,6) |
| Allowed Hello Loss | 2 |
| Net Diameter | 35 |
| Node Traversal Time (Seconds) | 0.04 |
| Route Error Rate Limit (pkts/sec) | 10 |
| Time Out Buffer | 2 |
| TTL Threshold | 7 |
| Addressing Mode | IPV4 |

In Normal Operation, the chosen best paths were found to be as shown in Figure 10. Under the wormhole attack of one malicious node, the path has changed as it is shown in Figure 11. We notice that all paths pass by malicious node 36.

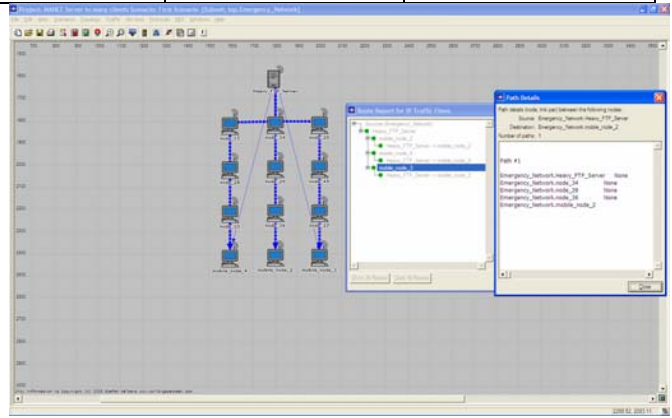

Figure 10. Chosen paths for the different demands in normal network operation for the first scenario

From the route report for IP traffic, the paths for each demand can be recorded as it is shown in Figure 11. These routes can be used in a program that calculates the number of routes traversing a specific node. If a node or a group of nodes were found to have more than a threshold number of routes passing by them, they become suspicious and further analysis could be done.

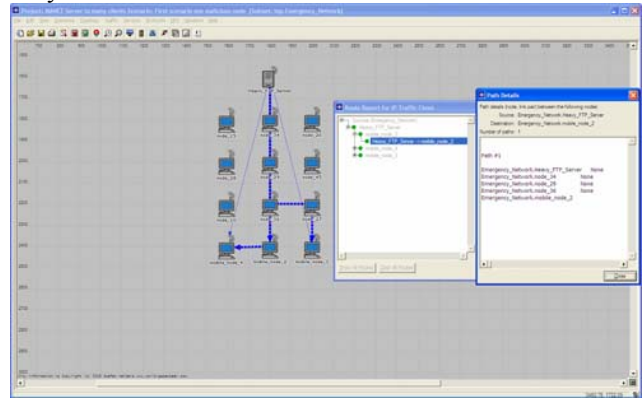

Figure 11. Chosen paths for different demands under the wormhole attack using one malicious node

Other different scenarios were simulated for the out of band channel wormhole attack, using two malicious nodes.

Results have shown that, not only the traffic passes via the malicious nodes, but also intermediate nodes have been bypassed and hence the chosen path appears to have less number of hops.

Figure 12 shows the ad hoc network routing under normal operation, and Figure 13 shows the wormhole attack with the two malicious nodes node 10 and node 15. From the path details screen shown in Figure 13, we notice that the routing does not pass by the node in between (node 28). To prove that the path chosen under wormhole attack should have normally passed through node 28, we have forced the routing to go through this path. As it is shown in Figure 14, node 28 is included in the path.

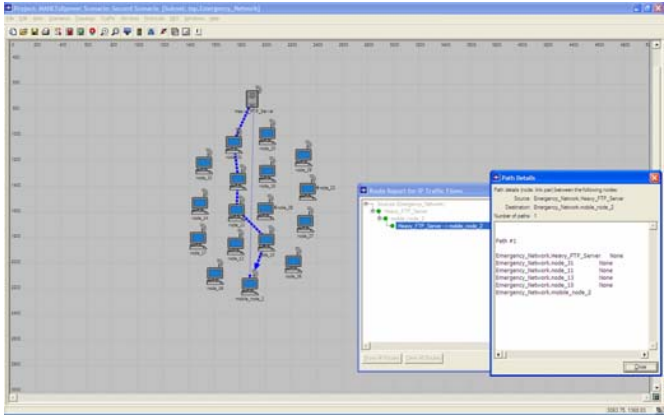

Figure 12 Chosen path for the configured demand in normal network operation for the second scenario

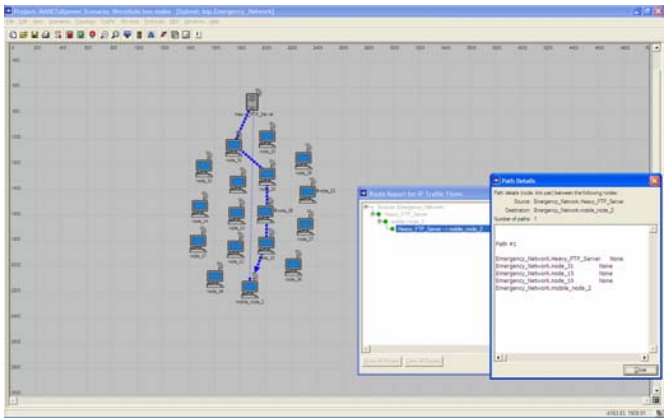

Figure 13 Ad hoc network under the wormhole attack of two colluding nodes

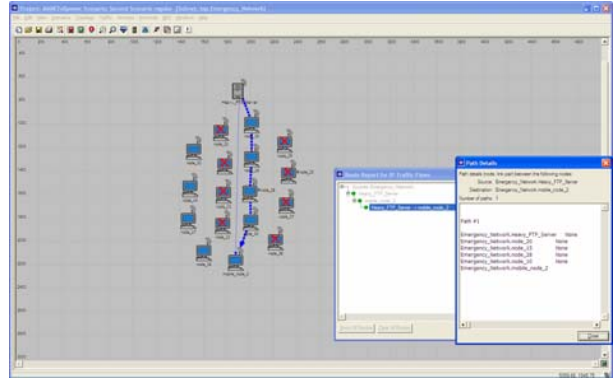

Figure 14 Regular forced same wormhole path without the wormhole attack for the second scenario

Finally, we present another scenario for the wormhole attack in the presence of three malicious nodes in a network consisting of 35 nodes with the network parameters as in Table III.

TABLE III. AD HOC NETWORK PARAMETERS FOR THE THIRD SCENARIO

| Ad Hoc Network Parameters | Values |
|---|---|
| Route Request Retries | 5 |
| Route Request Rate (pkts/sec) | 10 |
| Active Route Time Out (Seconds) | 3 s |
| Hello Interval (Seconds) | Uniform (1,1.1) |
| Allowed Hello Loss | 10 |
| Net Diameter | 35 |
| Node Traversal Time (Seconds) | 0.04 |
| Route Error Rate Limit (pkts/sec) | 10 |
| Time Out Buffer | 2 |
| TTL Threshold | 10 |
| Addressing Mode | IPV4 |

Figure 15 illustrates the normal network routing for a configured demand between the source and destination nodes. Figure 16 shows the routing under the wormhole attack in the presence of three malicious nodes: node 12, node 14, and node 31. Finally, Figure 17 shows the forced routing in the same wormhole path without the presence of malicious nodes to show the effect of the wormhole attack bypassing node 13 and node 15.

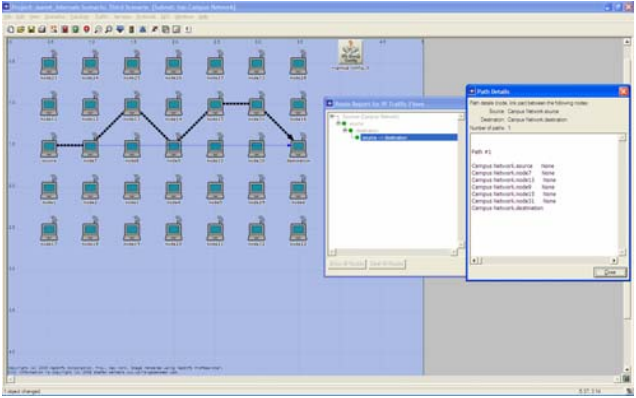

Figure 15. Normal routing in the third scenario

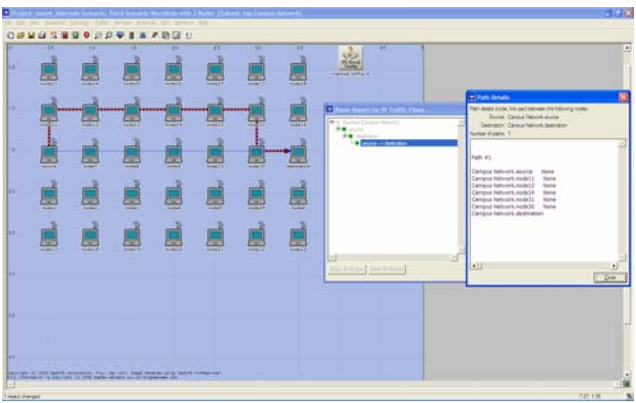

Figure 16. Selected path under the wormhole attack using three malicious nodes

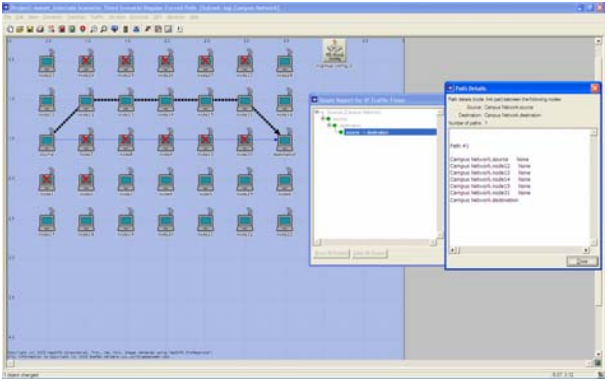

Figure 17. Forced regular path in the absence of the wormhole attack

## V. COMPLEX WORMHOLE ATTACKS

In section III, we analyzed each attack mode in details and pointed to its added benefits from an attacker's point of view. In this section, we suggest the improvement of wormhole attacks, from an attacker's perspective by introducing the concept of complex wormhole attacks. By complex wormhole attacks we mean the merging of two or more modes of wormhole attacks to benefit from their added advantages.

The first complex attack uses encapsulation in combination with directed out of band channel. This can be done by encapsulating the RREQ packets, sending them to colluding node using a long-range directional wireless link or a direct wired link. The benefits of this complex attack are having a faster and more guaranteed arrival of RREQs. The challenge in this method is to find a colluding node close enough to destination, a good distribution and communication of colluding nodes is needed. To overcome this challenge, a list of neighbors could be exchanged between colluding nodes, or more practically we can just encapsulate the RREQ and send it to all colluding nodes. One could think that another complex attack might be launched using encapsulation in conjunction with high power transmission as in that case the encapsulated packet will arrive to ANY node which is more dangerous and meaningless from the attacker's point of view!

Encapsulation can also be used jointly with protocol deviations to send the packet faster however the encapsulation itself will take more time. The best scenario is to have a large number of malicious nodes that do not back off, now for this attack the challenge is to check for the minimum number of malicious nodes that would be effective in launching such an attack. Thus those two complex attacks have used encapsulation, and to compensate the added time of encapsulation, other attack modes were used in conjunction to speed up the attack and make it more efficient. To hide the malicious node identities during encapsulation, packet relay can be used. The most important is to have a predetermined clear route between colluding nodes. In that case packet relay mode has also benefited from sending regular packets and not RREQ packets within the network, which makes the hiding of packets much easier.

If there is a suitable number of malicious nodes on the path between two nodes out of range that manage to send their control packets to convince them they are neighbors, using the packet relay mode of attacks, the use of protocol deviation will add a speed to the delivery of control packets, which makes the path more attractive from the points of view speed and number of hops as well.

For the purpose of speeding up the packet delivery while hiding the malicious nodes identities, packet relay could be used with out of band channels, this will help to minimize the needed number of relay nodes to convince two extremely distant nodes that they are neighbors. This is also much safer for the attacker, while having more attractive speedy paths with minimum number of hops. However this channel, as we mentioned before, needs special arrangements. Packet relay could also be used with high power transmission this will add the two important features of both modes which are anonymity and speed. More complex attacks could be launched using three attack modes in conjunction. For instance, encapsulation with packet relay and protocol deviations. In that case encapsulation is used to transform control packets to regular packets, packet relaying is used to hide the identities of encapsulating nodes and making them invisible, and finally either protocol deviation could be used to speed up the packets routing. Out of band channels could also be used as the third stage, to make the attack speedier, however they suffer from the need of special arrangements.

## VI. CONCLUSIONS AND FUTIRE WORK

In this paper we introduced the wormhole attack, presented its different modes in details together with an attack graph that we constructed to illustrate the sequence of events in each mode. We also discussed the threats that this attack presents briefly, and overviewed the effort done in the literature to combat this attack. While wearing an attacker's hat, we analyzed each mode and identified its advantages, disadvantages, challenges, possible solutions to these challenges, minimum number of nodes to launch each attack mode, suitable network topology, and countermeasures that could be used and have to be considered while launching each wormhole attack mode. To illustrate this attack's effect we presented the simulation results of two modes of this attack. The analysis has helped us, while still wearing the attacker's hat, to improve the wormhole attack by introducing the concept of complex wormhole attacks. In this type of attacks many modes have been suggested to be used in conjunction to benefit from the advantages of each to compensate for other modes disadvantages. Ethically, this type of wormhole analysis is important to account for possible new dangers and variations of this attack. Furthermore, it can help in putting some constraints on the network topology to design a robust network for such attacks, and in the design of new and more powerful attack countermeasures. In the future we plan to simulate complex attacks and compare their performance to select the optimum complex attack method from an attacker's point of view. Once selected, it will be tested with some of the proposed countermeasures and will help in the development of new attack prevention and detection schemes.